\documentclass[aps,pre,twocolumn,floatfix,showpacs,superscriptaddress]{revtex4}

\usepackage{graphicx}
\usepackage{bbm}
\usepackage{amsmath,amssymb}
\usepackage{color,calc}

\newcommand{\be}{\begin{equation}}
\newcommand{\beq}{\begin{equation}}
\newcommand{\ee}{\end{equation}}
\newcommand{\bea}{\begin{eqnarray}}
\newcommand{\eea}{\end{eqnarray}}
\newcommand{\ba}{\begin{array}}
\newcommand{\ea}{\end{array}}

\newif\ifprintcomments

\printcommentstrue  % whether to print the comment boxes or not.

\begin{document}
\title{Billiards in magnetic fields: A molecular dynamics approach}

\author{M. Aichinger}
\affiliation{Johann Radon Institute for Computational and Applied
Mathematics (RICAM), Austrian Academy of Sciences,  Altenberger Strasse 69, A-4040 Linz, Austria}

\author{S. Janecek}
\affiliation{Institut f\"ur Theoretische Physik, Johannes Kepler 
Universit\"at, A-4040 Linz, Austria}
\affiliation{Institut de Ci\`encia de Materials de Barcelona, Campus
  UAB, 08193 Bellaterra, Barcelona, Espa\~na.}

\author{E. R{\"a}s{\"a}nen}
\email[Electronic address:\;]{erasanen@jyu.fi}
\affiliation{Nanoscience Center, Department of Physics, 
University of Jyv{\"a}skyl{\"a}, FI-40014 Jyv{\"a}skyl{\"a}, Finland}
\affiliation{Institut f\"ur Theoretische Physik, Johannes Kepler 
Universit\"at, A-4040 Linz, Austria}

\date{\today}


\begin{abstract}
  We present a computational scheme based on classical molecular
  dynamics to study chaotic billiards in static external magnetic
  fields. The method allows to treat arbitrary geometries and several
  interacting particles. We test the scheme for rectangular
  single-particle billiards in magnetic fields and find a sequence of
  regularity islands at integer aspect ratios.  In the case of two
  Coulomb-interacting particles the dynamics is dominated by chaotic
  behavior. However, signatures of quasiperiodicity can be identified
  at weak interactions, as well as regular trajectories at strong
  magnetic fields. Our scheme provides a promising tool to monitor the
  classical limit of many-electron semiconductor nanostructures and
  transport systems up to high magnetic fields.
\end{abstract}

\maketitle

\section{Introduction}

Classical and quantum billiard systems~\cite{gutzwiller,stockmann} are
of significant interest both in nonlinear physics and in applications
based on low-dimensional nanostructures~\cite{nakamura}. For example,
quasi-two-dimensional (quasi-2D) quantum dots~\cite{qd_review} are
studied in view of emerging applications in the field of quantum
computation~\cite{hanson}. They exhibit deterministic ballistic motion
of the electrons as "billiard balls" and provide the possibility to
tune their shape, size, and electron number. A particularly intriguing
feature is the connection between classical dynamics and the
statistical properties of the corresponding quantum system
\cite{berrytabor, bohigas}. For systems with mixed chaotic and regular
dynamics, the Berry-Robnik formula \cite{berryrobnik84} links the volume
ratio of regular and chaotic regions in classical phase space to the
quantum-mechanical level distribution \cite{makino99, makino01}.

External magnetic fields pose, on the one hand, an interesting
complication to classical (and quantum) billiards~\cite{robnik}, and,
on the other hand, provide an easily accessible way to experimentally
{\em control} the particle dynamics. Recently,
magnetic fields have been used to manipulate electron transport in
coupled electron billiards~\cite{brunner}. In many cases, e.g., in
rectangular~\cite{berglund,aguiar,date,narewich} or triangular~\cite{christensson}
billiards, an external
magnetic field leads to mixed dynamics between regularity and
chaoticity. The breaking of time reversal symmetry due to the presence
of a magnetic field results in new properties of the level spacing
statistics of the corresponding quantum system \cite{berryrobnik86,
  robnikberry86}. 

In contrast to freely tunable parameters, such as external magnetic
and electric fields, {\em interactions} between particles are {\em
 inevitably} present in any realistic physical system. While
single-particle billiards have been studied thoroughly for many years
now, billiards of \emph{interacting} particles are still a relatively
young field. Classical billiards for two interacting particles
have been studied using various models, e.g., Coulomb-like
interactions in a one-dimensional box~\cite{lilia} and in an
isotropic~\cite{aaberg} and anisotropic harmonic
oscillator~\cite{drouvelis}, as well as applying hard-sphere contact
interaction in a rectangular~\cite{awazu} and a mushroom-shaped
box~\cite{lansel}. The statistical mechanics of such systems has also
been extensively studied recently~\cite{statmech}.  Quantum-mechanically,
interaction-induced chaos has been studied in a two-electron quantum
dot~\cite{aaberg,drouvelis,n2_quantum}, and, very recently, also in
the framework of time-dependent density-functional
theory~\cite{adam,tddft} -- an approach that might enable examination
of quantum chaos in systems containing a large number of interacting
particles.

Single-particle billiards have traditionally been studied by either
reducing the dynamics of the system to a bouncing map (for magnetic
single-particle billiards, see, e.g., \cite{meplan}), or by
investigating the infinitesimal variations of the trajectories using
the method of Jacobi fields \cite{voros, tasnadi}. In an interacting
billiard, however, the trajectory of a particle between successive
bounces is not known in advance, as its motion is coupled to the
motion of all other particles. The locations of the bounces at the
wall are not given by simple geometric considerations anymore, and
thus the methods used to study single-particle systems do not carry
over in a straightforward way.

In this paper, we present a classical molecular dynamics scheme that
allows to calculate the trajectories of interacting particles in an
arbitrary 2D billiard system exposed to a uniform and perpendicular
magnetic field. To demonstrate the method, we focus on single- and
two-particle dynamics in rectangular billiards. In the
single-particle case, we present an efficient method to systematically
obtain ``regular'' and ``chaotic'' regions in phase space, which
allows us to monitor the combined effect of the magnetic field and the
rectangle shape. We find a pattern of increased regularity at integer
aspect ratios. In the two-particle case mostly chaotic behavior is
found, but also regular orbits at high magnetic fields. The relevance
of the method in studying the classical limit of collective effects in
many-electron structures is discussed.

\section{Method}

\subsection{Propagation of particles}\label{moldyn}

To calculate the trajectories of charged particles, we use a modified
velocity verlet algorithm suited for incorporating arbitrarily strong
static homogeneous external magnetic fields~\cite{spreiter:1999}.
With a magnetic field $\mathbf{B}=(0,0,B)$ pointing in {\em z}
direction, the acceleration of a charged particle reads
\begin{equation}
\mathbf{a}(t)\,=\,\mathbf{a}^C(t)-\Omega\,\mathbf{e}_z\times\mathbf{v}(t),
\end{equation}
where $\mathbf{a}^C(t)$ is the velocity-independent part of the
acceleration depending only on external forces, and $\Omega=qB/m$ is
the cyclotron frequency for a particle with charge $q$ and mass
$m$. We use Hartree atomic units throughout the paper, such that
$\hbar=e=m_e=1/(4\pi\epsilon_0)=1$ and the velocity of light has the
value $c\approx 137.036$. Furthermore, the factor $1/c$ in the Lorentz
force law is absorbed into $B$, such that we have $\Omega=B$ for
electrons.  Within the modified velocity verlet algorithm presented in
Ref.~\cite{spreiter:1999}, each particle is propagated using the
following equations:
\begin{widetext}
\begin{equation}
  \begin{split}
    r_x(t+\Delta t)\,= &\,r_x(t)+\frac{1}{\Omega} \Big[ v_x(t)\sin(\Omega\Delta t)-
    v_y(t)\left[\cos(\Omega\Delta t)-1\right]\Big]+\\
    +&\frac{1}{\Omega^2}\Big[-a_x^C(t)\left[\cos(\Omega\Delta t)-1\right] -a_y^C(t) \left[ \sin(\Omega\Delta t)-
    \Omega\Delta t\right]\Big]+
    \mathcal{O}\left[(\Delta t)^3\right]\label{eq:rx}
\end{split}
\end{equation}
\begin{equation}
  \begin{split}
    r_y(t+\Delta t)\,=& \,r_y(t)-\frac{1}{\Omega} \Big[ - v_y(t)\sin(\Omega\Delta t)-v_x(t)
    \left[ \cos(\Omega\Delta t)-1\right] \Big]+\\
    +&\frac{1}{\Omega^2} \Big[ -a_y^C(t)\left[\cos(\Omega\Delta t)-1 \right]- 
    a_x^C(t)\left[-\sin(\Omega\Delta t) +\Omega\Delta t \right] \Big]+
    \mathcal{O}\left[(\Delta t)^3\right]
  \end{split}
\end{equation}
%%%
\begin{equation}
\mathbf{a}^C(t+\Delta t)\,=\,\mathbf{a}^C\left[r_1(t+\Delta t),\dots,r_N(t+\Delta t);t+\Delta t\right]
\end{equation}
%%%
\begin{equation}
  \begin{split}
    v_x(t+\Delta t)\,= &\,v_x(t)\cos(\Omega\Delta t)+v_y(t)\sin(\Omega\Delta t)+
    \frac{1}{\Omega} \Big[ -a_y^C(t)\left[\cos(\Omega\Delta t)-1\right]+\\
    +&a_x^C(t)\sin(\Omega\Delta t) \Big]+
    \frac{1}{\Omega^2}\left[-\frac{a_x^C(t+\Delta t)-
        a_x^C(t)}{\Delta t }\left[\cos(\Omega\Delta t)-1\right] - \right.\\
     -&\left.\frac{a_y^C(t+\Delta t)-a_y^C(t)}{\Delta
         t}\left[\sin(\Omega\Delta t)-\Omega\Delta t\right]\right]
    +\mathcal{O}\left[(\Delta t)^3 \right]  
  \end{split}
\end{equation}
\begin{equation}
  \begin{split}
    v_y(t+\Delta t)\,= &\,v_y(t)\cos(\Omega\Delta t)-v_x(t)\sin(\Omega\Delta t)-
    \frac{1}{\Omega}\Big[ -a_x^C(t)\left[\cos(\Omega\Delta t)-1\right]-\\
    -&a_y^C(t)\sin(\Omega\Delta
    t)\Big]+\frac{1}{\Omega^2}\left[-\frac{a_y^C(t+\Delta
        t)-a_y^C(t)}{\Delta t} \left[\cos(\Omega\Delta t)-1\right]\right.-\\
    -& \left.-\frac{a_x^C(t+\Delta t)-a_x^C(t)}{\Delta t} \left[
        -\sin(\Omega\Delta t)+ \Omega\Delta t\right]\right] +
    \mathcal{O}\left[(\Delta t)^3\right] \label{eq:vy}
\end{split}
\end{equation}
\end{widetext}

\subsection{Phase space maps for single-particle billiards}\label{phase_space}

In a single-particle billiard system, the kinetic energy, and
consequently also the velocity $v=(v_x^2 + v_y^2)^{1/2}$ of the
particle, is a constant of motion. The dynamics of the billiard is
determined by the boundary conditions (see below) and the relative
strength of the magnetic field. The latter quantity is here given by
a parameter
\begin{equation}\label{eq:mu}
  \mu = R_c/L_x,
\end{equation}
where $R_c = v/B$ is the cyclotron radius and $L_x$ is the length 
of one side of the system (here a rectangle). 
The constant of motion can be used to reduce the four-dimensional phase
space $(x,y,v_x,v_y)$ to a three-dimensional (3D) one, where we have
chosen the
space spanned by $(x,y,v_x)$. To identify regular and chaotic regions
in this phase space, we use the following procedure:
\begin{enumerate}
\item We choose a 2D cross section ($x,v_x$) through the 3D phase space
  and divide it into a number of cells.
\item For one cell, we pick two phase space points in the cell that
  are very close to each other, but not identical up to the numerical precision.
\item We follow the trajectories through these two points for a
  certain propagation distance $s_\mathrm{tot}$ using 
Eqs.~(\ref{eq:rx})--(\ref{eq:vy}), and record all cells in the cross
  section through which they pass.
\item After having propagated for a distance $s_\mathrm{tot}$, we
  calculate the distance between the points in phase space, which is a
  measure of the ``regularity'' of the trajectory. We save this distance to
  all cells we have passed. If a cell has already been passed by a
  previous run, we take the maximum of the distances.
\item We start over from point (2) by picking another cell that has not yet
  been traversed by a trajectory, and repeat the whole process until
  all cells have been hit by a trajectory at least once.
\item We then plot the distances stored in the cells of our 2D cross
  section as a color-coded ``matrix plot''. In the following, these
  plots will be called ``phase-space maps''. Small numbers correspond
  to ``regular'' phase space cells, large numbers to ``chaotic''
  cells (see below for details).
\end{enumerate}

The algorithm can be efficiently parallelized, because trajectories
originating from different cells can be propagated independent of each
other. Our code uses the Message Passing Interface (MPI) and a
master/slave paradigm. The master process keeps track of
the phase space map and distributes free cells, i.e., cells
that have not yet been hit by any trajectory) to the workers. The
workers perform the propagation of the trajectories and communicate
the traversed cells and the phase space distance after the
propagation distance $s_\mathrm{tot}$ back to the server.

\section{Results}\label{results}

\subsection{Single particle}

We demonstrate our computational scheme by considering rectangular 
billiards with side lengths $L_x=1$ (fixed) and $L_y=\beta\,L_x$ 
(varied), where $\beta$ is the aspect ratio. 
The strength of the external magnetic field has been fixed to $B=1$,
so that the cyclotron radius $R_c=v/B$ is determined by varying the
velocity of the particle. In the single-particle case,
we focus on the dynamics of the system as a function of $\beta$ and 
$\mu=R_c/L_x$. In both of the limits $\mu\rightarrow 0$ and 
$\mu\rightarrow \infty$ the motion is regular, the former
corresponding to infinitely many circular orbits (cf. Landau-level
condensation in confined quantum systems) and the latter corresponding 
to linear
motion at zero field, which is always regular in rectangular billiards.
At $0<\mu<\infty$ the dynamics is generally mixed except at 
particular values of $\mu$ when the system is completely 
chaotic~\cite{berglund}.

\begin{figure}
\includegraphics[width=0.99\columnwidth]{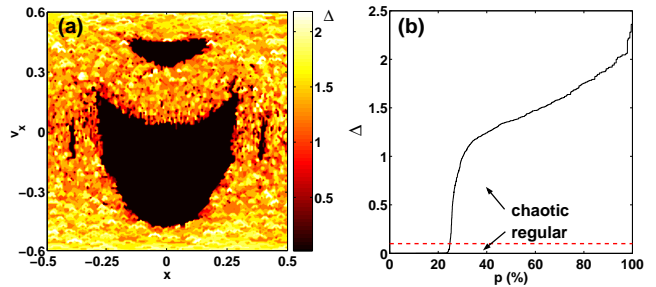}
\caption{(color online). (a) Phase space map for a rectangular billiard
with aspect ratio $\beta=2$. The color map indicates the
phase-space distance $\Delta$ between two orbits having a small initial
perturbation. (b) Ordered phase-space distances for all the
cells plotted in (a). The dashed line shows the threshold ($\Delta=0.1$)
between chaotic and regular motion.
}
\label{map}
\end{figure}

Fig.~\ref{map}(a)
shows an example of a phase space map calculated for the parameters
$\beta=2$ and $\mu=0.6$. The scheme described in
Sec.~\ref{phase_space} has been used to calculate the figure. The
cross section through the phase space has been partitioned into 150 cells
in each direction ($x$ and $v_x$). The color scale indicates the
phase-space distance $\Delta$ after propagating the trajectories by a
distance of $s_\text{tot} = 60\, L_x$. 

We find distinct areas of regularity associated with KAM
(Kolmogorov-Arnold-Moser) islands~\cite{gutzwiller}.
Figure~\ref{map}(b) shows the phase-space distances of all cells
sorted in ascending order. The sharp onset of the curve indicates a
distinct separation between regular and chaotic motion. To
consistently determine this separation, we choose a threshold of
$\Delta=0.1$ shown in the figure as a dashed line.  Thereby, this
particular system is regular by a fraction of $25\%$.  We presume that
by using a very high resolution it should be possible to determine and
categorize phase-space cells corresponding to {\em weak
  chaos}~\cite{weak_chaos}.  This topic is, however, beyond the scope
of this work and left for future research.

\begin{figure}
\includegraphics[width=0.7\columnwidth]{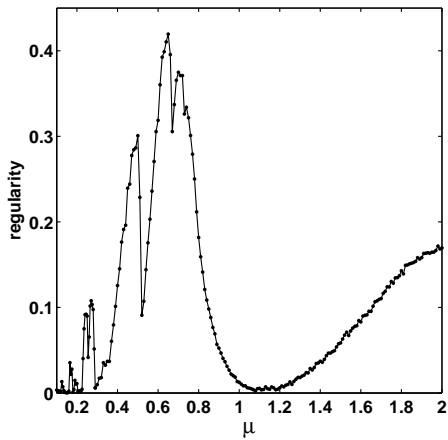}
\caption{Proportions of regularity in square billiards as a 
function of $\mu=R_c/L_x$, i.e., the ratio between the cyclotron
radius and the side length.
}
\label{square}
\end{figure}

In Fig.~\ref{square}
we show the proportions of regularity, estimated as shown in the
example in Fig.~\ref{map}, for square ($\beta=1$) billiards as a
function of $\mu$. We find excellent agreement with the result of
Berglund and Kunz~\cite{berglund} that has been calculated using an
exact method. This confirms the accuracy of the proposed scheme up to
strongly curvilinear motion, i.e., small values of $\mu$. Hence, we
expect the method to be reliable also in more complicated systems with many
particles and/or different boundaries.

\begin{figure}
\includegraphics[width=0.99\columnwidth]{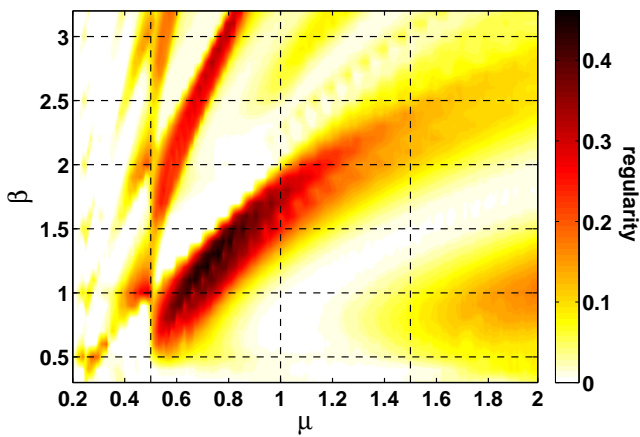}
\caption{(color online) Proportions of regularity
in rectangular billiards as a function of $\mu$ and
the aspect ratio $\beta$.
}
\label{recta}
\end{figure}

To assess the effect of the billiard shape onto the dynamics,
we have calculated the proportions of regularity as
a function of both $\mu$ and the aspect ratio $\beta$.
The result is shown in Fig.~\ref{recta}
for $0.3<\beta<3.2$ in steps of $0.1$ and for $0.2<\mu<2$ in steps of
$0.01$. Fig.~\ref{recta} required the calculation of 5430 phase space
maps, each consisting of 22 500 cells, thereby demonstrating the
numerical efficiency of the scheme.  Note that Fig.~\ref{recta} is not
symmetric around $\beta=1$, because we have varied
$L_y=\beta$ and thus the system {\em area} is not kept constant.

We find several islands of increased regularity centered at
$\beta=n/2$ with $n=1,2,3,\ldots$. Overall, the ``most regular'' case
is the square billiard ($\beta=1$), as expected. A more detailed analysis of
the regularity patterns and their connections to the periodic orbits
will be performed elsewhere.

\subsection{Two particles}\label{n2}

We now turn to the dynamics of two particles interacting
via Coulomb repulsion in a square well ($\beta=1$).
Now the velocities (and thus also the
cyclotron radii) are no longer constants of motion. 
The phase space is eight-dimensional, and
instead of the phase space map described in section \ref{phase_space},
we calculate so-called bouncing maps by recording the values
$(x,v_x)$ corresponding to the bounces of one of the particles
on the lower boundary ($y=0$) of the system.

We investigate the dynamics with different values for the ratio
\begin{equation}
  \label{eq:gamma}
  \gamma = \frac{E_{\rm k}(t=0)}{E_{\rm p}(t=0)}
\end{equation}
for the initial configuration, where $E_p=|{\mathbf r}-{\mathbf
  r}'|^{-1}/2$ is the Coulomb potential energy and $E_{\rm k}=v^2/2$
is the kinetic energy. 
The quantity $\gamma$ essentially
determines how ``strongly interacting'' the system is, as it
fixes the \emph{average} ratio of $E_k(t)$ and $E_p(t)$ for the full
{\em time-dependent} system through the initial energy components.
In physical applications this ratio could be varied by changing
either the particle density or the system size. A well-known 
example of the limit where the potential energy dominates is the 
Wigner crystal~\cite{wigner} forming in the electron gas at low densities.

\begin{figure}
\includegraphics[width=0.6\columnwidth]{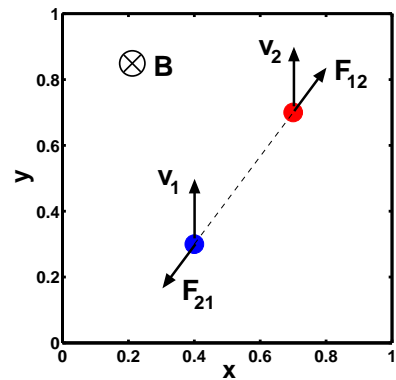}
\caption{(color online). Initial configuration of the calculations 
for two Coulomb-interacting particles in square billiards subjected
to a perpendicular magnetic field.
}
\label{initial}
\end{figure}

In the following examples we have
fixed the initial positions of the particles 
to $(x_1,y_1)=(2/5,3/10)$ and $(x_2,y_2)=(7/10,7/10)$.
The {\em initial} Coulomb energy in this case is $E_p(t=0)=2$.
After fixing $\gamma$ in Eq.~(\ref{eq:gamma}),
the initial kinetic energy $E_k(t=0)$
is distributed equally to both particles, and
the initial velocities ${\mathbf v}_1={\mathbf v}_2=(0,\sqrt{E_k})$ 
point in the {\em y} direction. The initial configuration is
visualized in Fig.~\ref{initial}.

The remaining parameter to be fixed is $\mu(t=0)$ defined in
Eq.~(\ref{eq:mu}). Note that again we fix only the initial
condition, and in the time-dependent run, the values of $\mu$ for both
particles vary due to
changes in the velocities. Since the initial velocity is determined
through $\gamma$, we fix $\mu(t=0)$ through $B$ in contrast with
the single-particle case where we always had $B=1$.

\begin{figure}
\includegraphics[width=0.7\columnwidth]{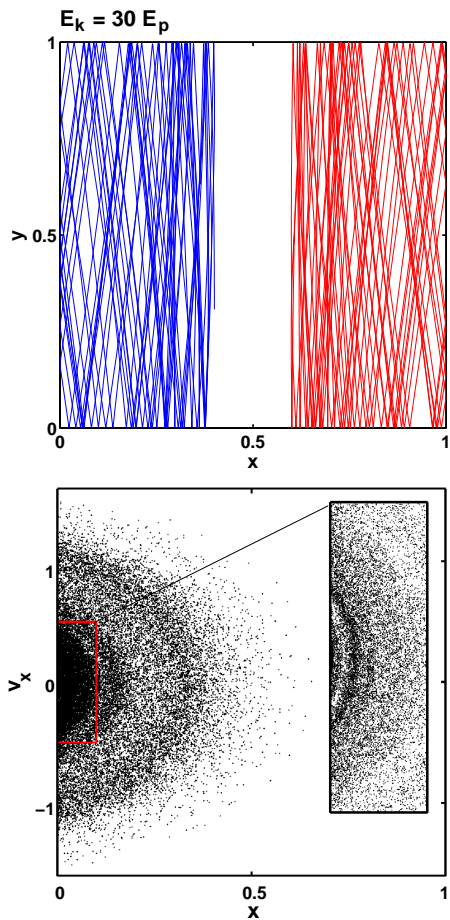}
\caption{(color online). Upper panel: Classical trajectories for 
two relatively weakly interacting particles 
indicated by blue (black) and red (gray) colors
in square billiards. The magnetic field is zero.
Lower panel: Bouncing map for the particle with 
the blue (black) trajectory in the upper panel.
}
\label{weak}
\end{figure}

First, we set $\gamma=30$ and the magnetic field to zero ($\mu\rightarrow\infty$)
and propagate sufficiently long to obtain a bouncing map with a large
number of points.
Figure~\ref{weak}
shows the trajectories of the particles up to $t=5$ (upper panel) and
the bouncing map up to $t=3\times 10^4$ (lower panel). The number of
bounces is $\sim 7.8\times 10^5$.  Apart from a few exceptions, the
particles remain separated in the left and right parts of the system
due to the Coulomb repulsion. However, as the interaction is
relatively weak, both particles move in the {\em y} direction, almost
undisturbed from their initial conditions. Close to the left and right
boundaries, where the interaction is weakest, the dynamics is most
regular. This can be seen in the trajectories, which are almost
straight lines in that regime. In addition, the bouncing map shows
regular curvilinear albeit blurry zones (see the inset in the lower
panel of Fig.~\ref{weak}).  These features may be designated as
quasi-regular motion in the system~\cite{weak_chaos}.

\begin{figure}
\includegraphics[width=0.7\columnwidth]{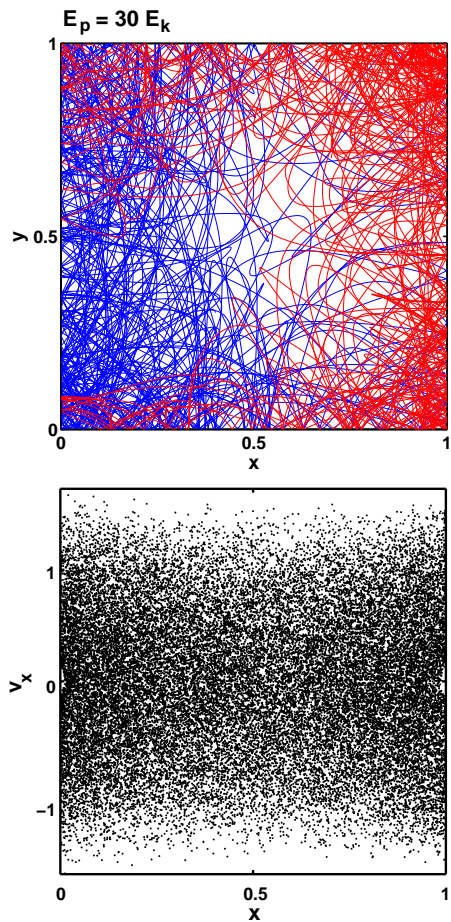}
\caption{(color online). Same as Fig.~\ref{weak} but for
relatively strongly interacting particles.
}
\label{strong}
\end{figure}

In the following example we keep the magnetic field at zero 
but increase the relative amount of interaction energy such
that $\gamma=1/30$. The trajectories and bouncing map
are shown in Fig.~\ref{strong}.
Here, the dynamics is very different from the weakly interacting 
case. Both particles occupy the whole area of the system, but due
to their strong repulsion, the corners are considerably more occupied 
than the central region, which is characterized by "scattering"
trajectories of high curvature. The bouncing map in the lower
panel is completely chaotic. Increasing the interaction even
further would enable to study classical Wigner crystallization~\cite{wigner}
in a dynamic picture. In the present system, for example, the
Wigner crystal would consist of two diagonal configurations summed
up to a four-point crystal.

\begin{figure}
\includegraphics[width=0.7\columnwidth]{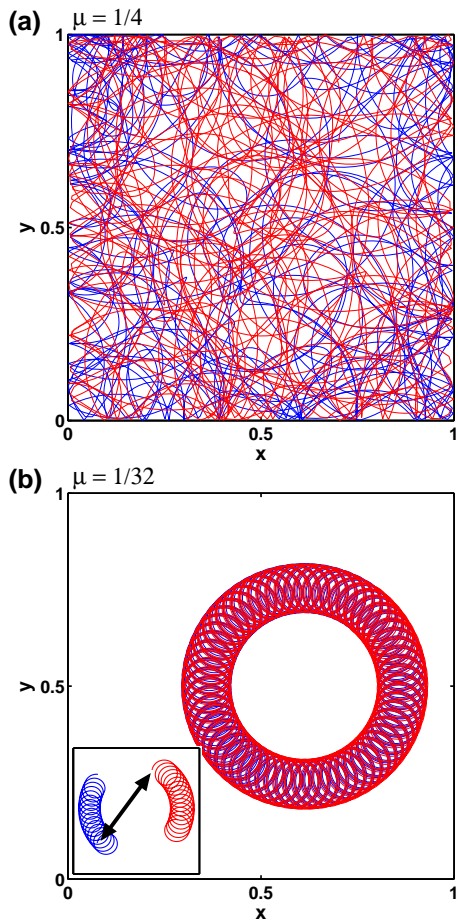}
\caption{(color online). Classical trajectories for 
two interacting particles at two different magnetic 
fields corresponding to $\mu=1/4$ and $\mu=1/32$, respectively.
}
\label{mag}
\end{figure}

Finally, we consider two systems with $\gamma=2$, 
where the magnetic field is set 
to values corresponding to $\mu(t=0)=1/4$ and $\mu(t=0)=1/32$, respectively. 
The trajectories are plotted in Fig.~\ref{mag}.
In the first case (a) the system seems to be fully chaotic, whereas
the latter configuration (b) leads to regular isolated orbits forming
a ring-like structure. In this case, the ``interaction axis'' (i.e.,
the dashed line in Fig.~\ref{initial}) performs a circular motion
that is superimposed by strongly confined cyclotron motions
at the opposite ends of the axis. The characteristics of this motion are
further illustrated in the inset of Fig.~\ref{mag}(b), which shows the
trajectories soon after the beginning of the time propagation. The
--- at a first glance --- counterintuitive result that a repulsive
interaction leads to bound motion can be understood by considering the
combined effect of Coulomb repulsion and the strong magnetic
confinement through the cyclotron motion. When the particles increase their
relative distance, the gain in kinetic energy
(at the expense of Coulomb energy) results in an increased radius of the cyclotron
motion. The different curvature of the trajectory on the different
sides of the ``circle'' gives rise to a bent cycloidal motion which can,
for the right choice of the parameters, lead to a bound motion as
depicted in Fig.~\ref{mag}(b).

The above results on the classical dynamics in magnetic fields suggest
to study the relation to the corresponding quantum mechanical
situation in semiconductor quantum dots. In fact, interesting vortex
patterns and edge localization have been found in rectangular
many-electron quantum dots at high magnetic fields~\cite{recta}. Such
patterns may exist -- in a statistical picture -- also in a classical
system.  A particularly interesting case would be the
quantum-mechanical analog of the bound motion shown in
Fig.~\ref{mag}(b).  Moreover, our scheme would allow to study the
effects of interactions on the classical limit of electron transport
in billiard arrays~\cite{brunner}.

\section{Summary}

We have introduced a computational scheme based on
molecular dynamics to study classical billiards of interacting 
particles in external magnetic fields. The accuracy and
efficiency of the method has been demonstrated in rectangular 
billiards. We have found excellent agreement with numerically
exact method in single-particle square billiards as a function
of the magnetic field. Changing the aspect ratio $\beta$ 
of the rectangle leads to islands of increased regularity 
at $\beta=n/2$ with $n=1,2,3,\ldots$. 
In square billiards of two interacting particles we have
found signatures of quasiperiodic orbits at weak interactions, and 
localization at strong interactions. Large magnetic
fields may lead to regular patterns also for
interacting particles. The scheme opens up the path
to study the classical limit of realistic many-particle systems 
related with, e.g., electronic transport experiments in 
mesoscopic structures.

\begin{acknowledgments}
We thank Roland Brunner for useful disdcussions. 
This work was supported by the Academy of Finland.
\end{acknowledgments}

\end{document}